\documentclass[final,5p,times,twocolumn]{elsarticle}
\usepackage{graphics}
\usepackage{amssymb}
\usepackage{amsmath}
\usepackage{colortbl}
\usepackage{xcolor}
\usepackage{textcomp}
\usepackage[utf8]{inputenc}
\usepackage{textcomp}
\usepackage{palatino}
\usepackage{lineno}

\DeclareMathOperator\erfi{erfi}

\journal{Nuclear Instruments and Methods in Physics Research A }

\begin{document}

\begin{frontmatter}

\title{Simulation design for forthcoming high quality plasma wakefield acceleration experiment in linear regime at SPARC\_LAB}

\author[lnf]{S. Romeo}\ead{stefano.romeo@lnf.infn.it}
\author[lnf]{E. Chiadroni}
\author[lnf,roma1]{M. Croia}
\author[lnf]{M. Ferrario}
\author[lnf]{A. Giribono}
\author[lnf]{A. Marocchino}
\author[roma1,SBAI]{F. Mira}
\author[lnf]{R. Pompili}
\author[infn-mi]{A.R. Rossi}
\author[lnf]{C. Vaccarezza}

\address[lnf]{Laboratori Nazionali di Frascati, Via Enrico Fermi 40, 00044 Frascati (Rome), Italy}
\address[infn-mi]{INFN-Milan and Department of Physics, University of Milan, Via Celoria 16, 20133 Milan, Italy}
\address[roma1]{University of Rome "Sapienza", Piazzale Aldo Moro 5, 00185 Rome, Italy}
\address[SBAI]{SBAI Department, University of Rome "Sapienza", Via Antonio Scarpa 14, 00161 Rome, Italy}

\begin{abstract}
In the context of plasma wakefield acceleration beam driven, we exploit a high density charge trailing bunch whose self-fields act by mitigating the energy spread increase via beam loading compensation, together with bunch self-contain operated by the self-consistent transverse field. The work, that will be experimentally tested in the SPARC\_LAB test facility, consists of a parametric scan that allows to find optimized parameters in order to preserve the high quality of the trailing bunch over the entire centimeters acceleration length, with a final energy spread increase of $0.1\%$ and an emittance increase of $5$ nm. The stability of trailing bunch parameters after acceleration is tested employing a systematic scan of the parameters of the bunches at the injection. The results show that the energy spread increase keeps lower than $1\%$ and the emittance increase is lower than $0.02$ mm mrad in all the simulations performed. The energy jitter is of the order of $5\%$.
\end{abstract}


\end{frontmatter}
   
\section{Introduction}

{
In this work we propose an ideal numerical study for a scheme of high quality plasma wakefield acceleration beam driven \citep{chen1987possible,esarey1996overview,hogan2005multi,blumenfeld2007energy,muggli2013physics,litos2014high} in the range of interest of the SPARC\_LAB test facility \citep{ferrario2013sparc_lab}. The study is ideal because the used bunches are generated in code and bi-Gaussian. The plasma is assumed uniform flat top, since we did not consider at this stage the effects of longitudinal ramps and it has been evaluated that the radial dependency of plasma density is negligible for the SPARC\_LAB experimental configuration \citep{marocchino2017experimental}. This working point exploits a high density charge trailing bunch whose self-fields act by mitigating the energy spread increase via beam loading compensation, together with bunch self-contain operated by the self-consistent transverse field. The energy spread is preserved exploiting the flattening of the field generated by longitudinal beam loading \citep{van1985improving,Katsouleas1987}. The idea of accelerating a high density trailing bunch on the wake of a low density bunch has been originally presented by the same authors in a PhD thesis work \citep{romeo2017} and the exploitation of the beam loading effect to preserve bunch quality has also been considered in the context of the AWAKE experiment at CERN \citep{olsen2017emittance}. This work starts from the design of a working point involving a low brightness \citep{marocchino2015study,marocchino2017elba} driving bunch that generates an accelerating wake in linear regime and a very high brightness trailing bunch with parameters that are compatible with the bunches produced at the SPARC\_LAB photo-injector using the hollow beam velocity bunching scheme \citep{pompili2016beam}. The scheme robustness is tested using parameter jitters that were measured in preparation of the plasma acceleration experiment. In our simulations, we also exploit the robustness of the proposed working point. We define as ‘robustness’ the ability of a given scenario to tolerate small perturbations without producing a significant variation in the phase space dilution, i.e. for the trailing bunch. Robustness is exploited verifying that any small variations do not lead to any instability growth that might significantly affect or the energy spread nor the emittance. Instead of linearly varying each parameter that can lead us to a large number of simulations, we prefer to leverage on the Latin Hypercube Sampling generally used in this scenarios but in the field of RF accelerator. The LHS randomly selects each parameter in a given range of tolerance by generating a number of virtual machines with simultaneous parameter variation. The choice of LHS is also dictated by the necessity to consider correlated effects and to demonstrate the stability where no parameter dominates the seeding of instability.

\begin{figure*}[t!]
     \centering
     \includegraphics [width=1.\linewidth]{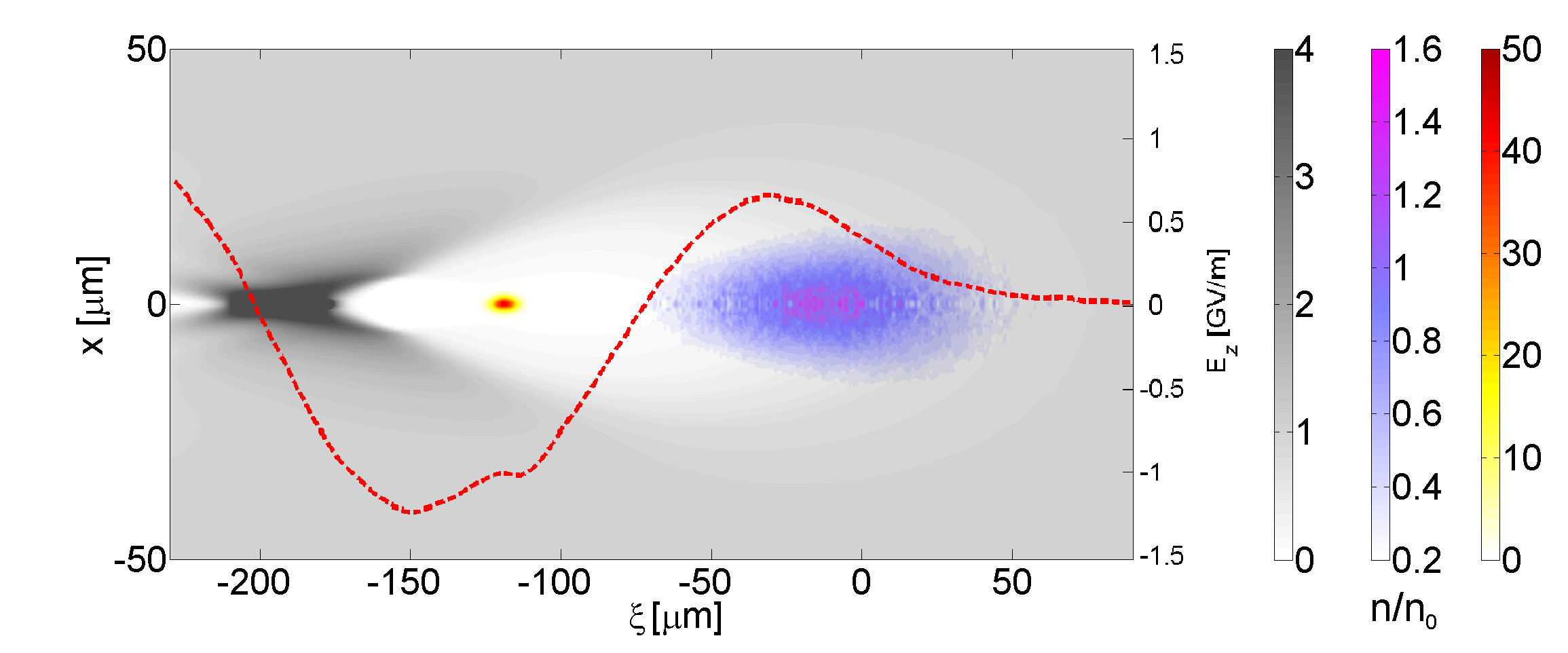}
     \caption{Accelerating scheme structure. As we can see the driver has a low density (purple scale, range [0.2 1.6]) and the witness has a very high density (orange scale, range [0 50]). The field is reported in red. We can notice the flattening in correspondence of the trailing bunch position.}
     \end{figure*}

The SPARC\_LAB photo-injector is composed of a 1.6 cell BNL/UCLA/SLAC type gun, operating at S-band ($2.856$ GHz) with a peak field of $120$ MV/m on the copper metallic photocathode that generates a $5.6$ MeV electron beam. The gun is then followed by two S-band and one C-band travelling wave (TW) sections whose accelerating gradient can boost the beam energy up to $160$ MeV, although in this work we will consider a scheme designed for bunches with $100$ MeV. The first S-band cavity is also used as RF compressor in velocity bunching regime ~\cite{serafini2001velocity}, setting the beam injection phase near to zero crossing. The first two S-band sections are surrounded by solenoid coils that can provide a magnetic focusing to better control the beam envelope and the emittance oscillations under RF compression.

\section{Simulation setup}

In order to achieve an accelerating gradient $>1$ GV/m with an energy increase of $\approx50\%$, we considered a plasma structure with a length of $5$cm with a background plasma density $2\times10^{16}$ [cm]$^{-3}$. The bunches are assumed bi-Gaussian and with cylindrical symmetry, with rms longitudinal and transverse dimension $\sigma_{z \ D}$ and $\sigma_{x,y \ D}$ for driving bunch and $\sigma_{z \ w}$ and $\sigma_{x,y \ w}$ for trailing bunch. We perform the parameters choice in the range of interest of the SPARC\_LAB photoinjector. The energy at the entrance was assumed $E$=102.2 MeV both for driving and trailing bunch. In order to maximize the accelerating gradient, the bunch length was chosen such that $k_p\sigma_{z \ D}=1$ with $k_p$ plasma wave number $k_p=\sqrt{e^2 n_0/(\epsilon_0 m_e c^2)}$ and $n_0$ background plasma density, that leads to $\sigma_{z \ D}=37.2$ µm. The transverse emittance of the driver is $\epsilon_{n \ D}=17$ mm mrad. The corresponding transverse matched spot size at the injection \citep{romeo2017} is

\begin{equation}
\sigma_{x,y \ D}=\sqrt{\dfrac{4}{\gamma \Lambda \mathcal{Z}}}\epsilon_n;
\end{equation}

where we used the longitudinal normalized plasma response function

\begin{equation}
\mathcal{Z}=\sqrt{\pi/2} \ k_p\sigma_{z \ D} \ e^{k_p^2\sigma_{z \ D}^2} \erfi (k_p\sigma_{z \ D}/\sqrt{2});
\end{equation}

the normalized bunch length $\Lambda=\alpha_D k_p^2 \sigma_{x,y \ D}^2$ and the normalized driving bunch density $\alpha_D=n_D/n_0$ with $n_D$ driving bunch density. Therefore, the injection spot size is $\sigma_{x,y \ D}=10.3$ µm with a corresponding normalized driving bunch density $\alpha_D=1$. In order to preserve the energy spread during the acceleration, the trailing bunch length is chosen very short $\sigma_{z \ w}=3$ µm. The transverse emittance of trailing bunch is set to $\epsilon_{n \ w}=0.3$ mm mrad. This corresponds to a matched spot size \citep{romeo2017}

\begin{equation}
\sigma_{x,y \ w}=\sqrt[4]{\dfrac{1}{\gamma}}\sqrt{\dfrac{2\epsilon_{n \ w}}{k_p}};
\end{equation}

that is $\sigma_{x,y \ w}=1.26$ µm. Initial energy spread of bunches are set to $\sigma_E=0.1\%$. For clarity, the driver and witness parameters are also listed in table 1, while in Fig.1 is shown the accelerating scheme.

\begin{table}[h!]
\caption {Beam parameters at the injection}
\begin{center}
\begin{tabular}{|c|c|c|}
\hline
&\cellcolor{green!10}Driver&\cellcolor{green!10}Witness\\
\hline
\cellcolor{green!10}$Q$ [pC] & 200& 10\\
\hline
\cellcolor{green!10}$\gamma$& \multicolumn{2}{c|}{$200$}\\
\hline
\cellcolor{green!10}$\epsilon_n$ [mm mrad] & $17$&$0.3$\\
\hline
\cellcolor{green!10}$\sigma_r$ [µm]  & $10.3$ & $1.26$\\
\hline
\cellcolor{green!10}$\sigma_z$ [µm] & $37.2$ &$3$ \\
\hline
\cellcolor{green!10}$\sigma_E$ [$\%$] & $0.1$ &$0.1$ \\
\hline
\end{tabular}
\end{center}
\end{table}

Simulations have been performed with the time explicit hybrid kinetic-fluid code $Architect$ \citep{marocchino2016efficient, massimo2016comparisons} Beams are treated kinetically in a 6D space with conventional particle in cell schemes, while the background plasma electrons are modeled as a cold relativistic fluid.\\
The electromagnetic fields that cause the motion are generated by the sum of the currents of the beams and the background plasma. Electromagnetic fields and fluid equations are solved in cylindrical symmetry with a moving window technique. The advantage of the hybrid nature is the capability to run a simulation in a few hours on a single core, this also allows for some systematic scan. The drawback is that the code, due to the fluid description of fields, poorely capture plasma wave-breaking at bubble closure. The code begins to lose accuracy for highly nonlinear regimes that can be mathematically identified for drivers whose parameters are $\alpha\gg1$ and $\tilde{Q}\gg 1$ ($\tilde{Q}=N_b k_p^3/n_0$ with $N_b$ the number bunch particles). In these case, the region of particle crossing is not well represented, despite the code keeps a good reliability on the description of the blow-out region where the trailing bunch is injected. A 2D description for the background and fields limit the possibility to run highly asymmetric effects, or to take into account for phenomenon such as the Hose instability. These effects are not relevant for our regimes and distances. The integration time step $\Delta t = 0.44$ fs. The mesh is squared with a dimension $0.75$ µm$\times0.75$ µm. The box is composed by a $732\hat{u}_r\times932\hat{u}_z$ grid cells, with $\hat{u}_z$ and $\hat{u}_r$ versors in the $z$ and $r$ dimension respectively, corresponding to a box dimension of $275$ µm $\hat{u}_r\times700$ µm $\hat{u}_z$. The driving bunch is located at $z=0$ and it is discretized with $4\times10^5$ particles. In the reference simulation the trailing bunch is located at $0.5\lambda_p$ ($\approx116.7$ µm) and discretized with $5\times10^4$ particles.

\section{Bunch separation scan}

The beam loading compensation of energy spread for a wake excited by low density driving bunch depends on the bunch separation \citep{Katsouleas1987}. Here we perform two scans in order to minimize the witness energy spread growth. In Fig.2 we report the outcoming energy spread versus the bunch separation at the injection. The energy spread is evaluated on the entire trailing bunch distribution and performing two different cuts. The cuts are performed on longitudinal phase spaces, neglecting the particles that give higher contribution to longitudinal emittance, in order to be consistent with the cuts that are performed experimentally \citep{filippetto2010high}. The two cuts are performed on $5\%$ (0.5 pC) and $10\%$ (1 pC) of charge respectively. The first scan is performed in the range $0.45\lambda_p \leq \Delta z \leq 0.55\lambda_p$ with a rough step of $0.05\lambda_p$. The second scan is performed in the range $0.5\lambda_p \leq \Delta z \leq 0.525\lambda_p$ with a fine step of $0.005\lambda_p$.

     \begin{figure}[h!]
     \centering
     \includegraphics [width=1.\linewidth]{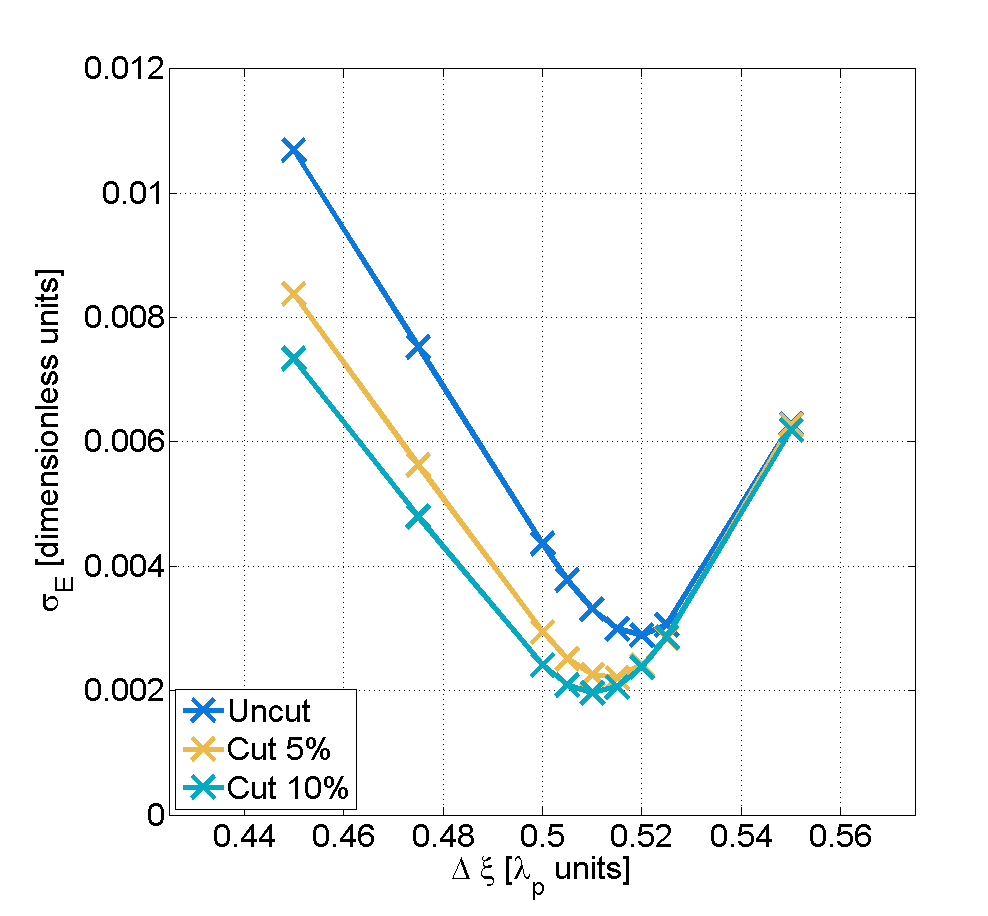}
     \caption{Final energy spread vs. bunch separation. The energy spread is evaluated on the entire particle distribution and with cuts operated on longitudinal phase space over $5\%$ and $10\%$ of charge.}
     \end{figure}     

The minimum energy spread depends on the cut performed. For the $10\%$ cut, the minimum energy spread is found for $\Delta \xi = 0.51\lambda_p$. A direct comparison of the cut longitudinal phase space for $\Delta \xi = 0.505\lambda_p$ (Fig.3) and $\Delta \xi = 0.51\lambda_p$ (Fig.4) show that the beam loading compensation of the core is more evident in the first case. The energy spread in the two cases differ for less than $0.01\%$, so we choose $\Delta \xi = 0.505\lambda_p$ as reference case. The trailing bunch parameters at the extraction are listed in table 2.

\begin{table}[h!]
\begin{center}
\caption {Witness parameters at injection and extraction}
\begin{tabular}{|c|c|c|}
\hline
&\cellcolor{green!10}Injection&\cellcolor{green!10}Extraction\\
\hline
\cellcolor{green!10}$\gamma$&$200$&$305$\\
\hline
\cellcolor{green!10}$\epsilon_n [$mm mrad$]$ & $0.3$&$0.305$\\
\hline
\cellcolor{green!10}$\sigma_r [{\mu}$m$]$ & $1.26$ & $1.1$\\
\hline
\cellcolor{green!10}$\sigma_z [{\mu}$m$]$ & $3$ &$3$ \\
\hline
\cellcolor{green!10}$\sigma_E [\%]$ & $0.1$ &$0.21$ \\
\hline
\end{tabular}
\end{center}
\end{table}

The average accelerating gradient is $1.07$ GV/m and the energy increase of the order of $50\%$. The energy spread increase is of $0.11\%$ and the emittance increase is of few nanometers. We notice that the quality of the trailing bunch is preserved despite a substantial energy increase.

     \begin{figure}[t!]
     \centering
     \includegraphics [width=1\linewidth]{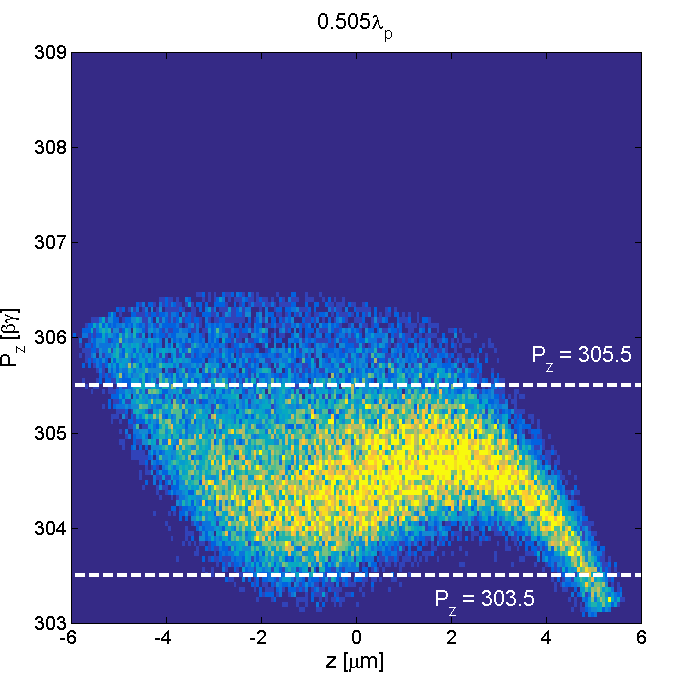}
     \caption{Longitudinal phase space of trailing bunch with a bunch separation of $\Delta \xi = 0.505\lambda_p$.}
     \end{figure}     

     \begin{figure}[t!]
     \centering
     \includegraphics [width=1\linewidth]{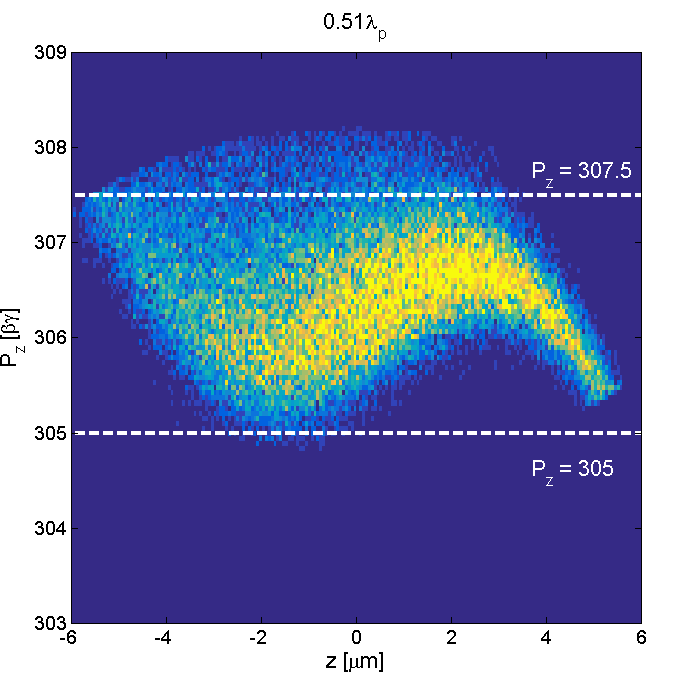}
     \caption{Longitudinal phase space of trailing bunch with a bunch separation of $\Delta \xi = 0.51\lambda_p$.}
     \end{figure}     
          
\section{Tolerance analysis}

In order to perform the tolerance analysis over the proposed working point we used the results of the measurements performed at the SPARC\_LAB photo-injector in order to evaluate the jitters of the bunch parameters at the plasma structure entrance. In the analysis we varied the following parameters at the injection

\begin{itemize}
\item[-] driver and witness transverse spot size $\pm15\%$;
\item[-] driver and witness length $\pm12\%$;
\item[-] driver and witness charge $\pm10\%$;
\item[-] bunch separation $\pm8\%$.
\end{itemize}

To further simplify the evaluations, we considered the same fluctuations in percentage for both driving and trailing bunch for any considered parameter. We neglected the expected correlation between the length of the bunches and bunch separation. Assuming these jitters we performed a simulation scan of the real photo-injector using the sampling approach of the latin hypercube (LHS) ~\cite{wyss1998user} with uniform jitter distribution.Each of the 30 simulations was designed combining a random variation of all parameters. The random variation of the single parameter has an uniform distribution. This approach was adopted since, respect to the variation of a single parameter, it considers also the correlated effects that can arise from the simultaneous variation of multiple parameters. The results are reported in Fig.5, Fig.6 and Fig.7. The accelerated beams result very stable in term of quality. The average energy spread on the sample is $\approx0.4\%$ with a standard deviation $\approx0.15\%$. The emittance growth is $<7\%$ in all the simulations considered. The most evident effect of jitter is related to the outcoming witness energy. The energy value is $E=156\pm7.2$ MeV corresponding to a fluctuation of $4.6\%$ on the average value.

\section{Summary}

We have designed and tested an ideal scheme for plasma wakefield acceleration beam driven experiment in the range of interest of the SPARC\_LAB facility. Through numerical scan we have shown that it is possible to obtain a working point that is suitable for real acceleration experiments. This working point was design performing a numerical scan involving the variation of a single parameter. The robustness of this scheme has been tested with very promising results. An important remark is that the ranges of the tolerance analysis were intentionally assumed higher than the average machine performance. Our analysis consider a wide enough number of parameters and scenarios that we expect a realistic case to be contemplated in our simplified numerical study. The analysis evidenced an high energy jitter of the trailing bunch at the extraction, that could cause a very difficult transport of the bunch downstream the plasma structure with all the relative consequences. Further evaluations of this aspect are required. Nevertheless, our conclusion is that this beam driven configuration guarantees an high range of parameters that allow an high quality acceleration.

\section*{Acknowledgments}

{This work was supported by the European Union's Horizon 2020 research and innovation programme under grant agreement No. 653782.}

     \begin{figure}[t!]
     \centering
     \includegraphics [width=.71\linewidth]{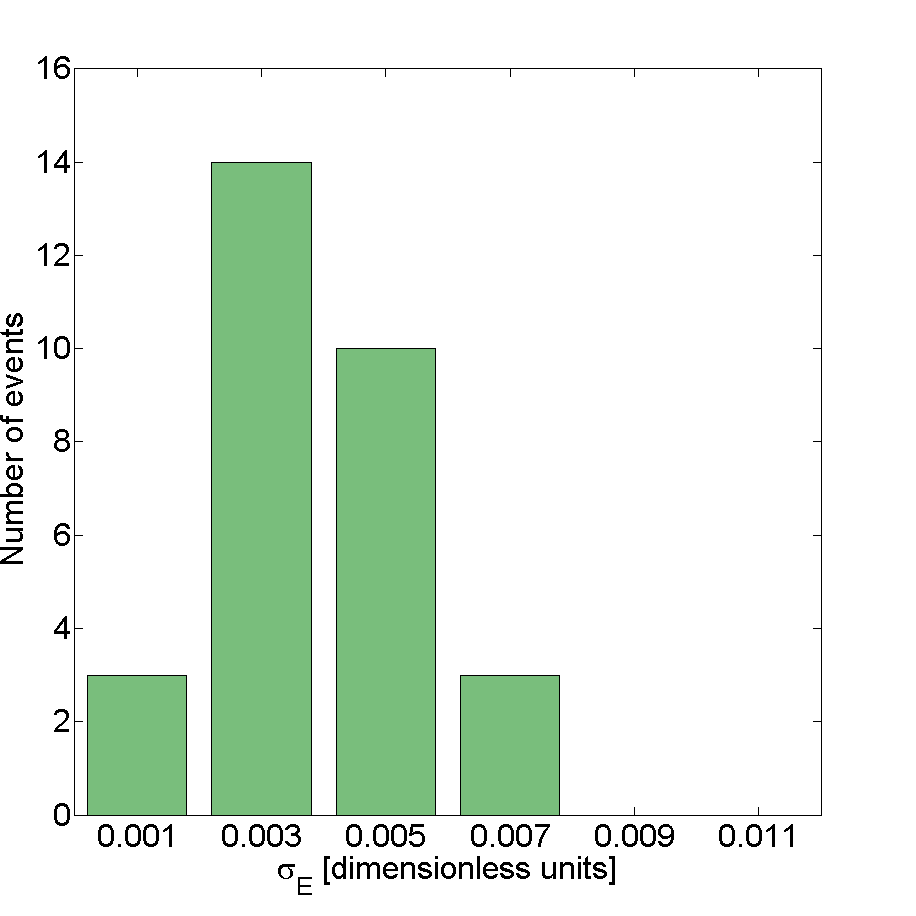}
     \caption{Energy spread jitter distribution of 30 machines in the latin hypercube sampling analysis performed for the considered working point.}
     \end{figure}

     \begin{figure}[t!]
     \centering
     \includegraphics [width=.71\linewidth]{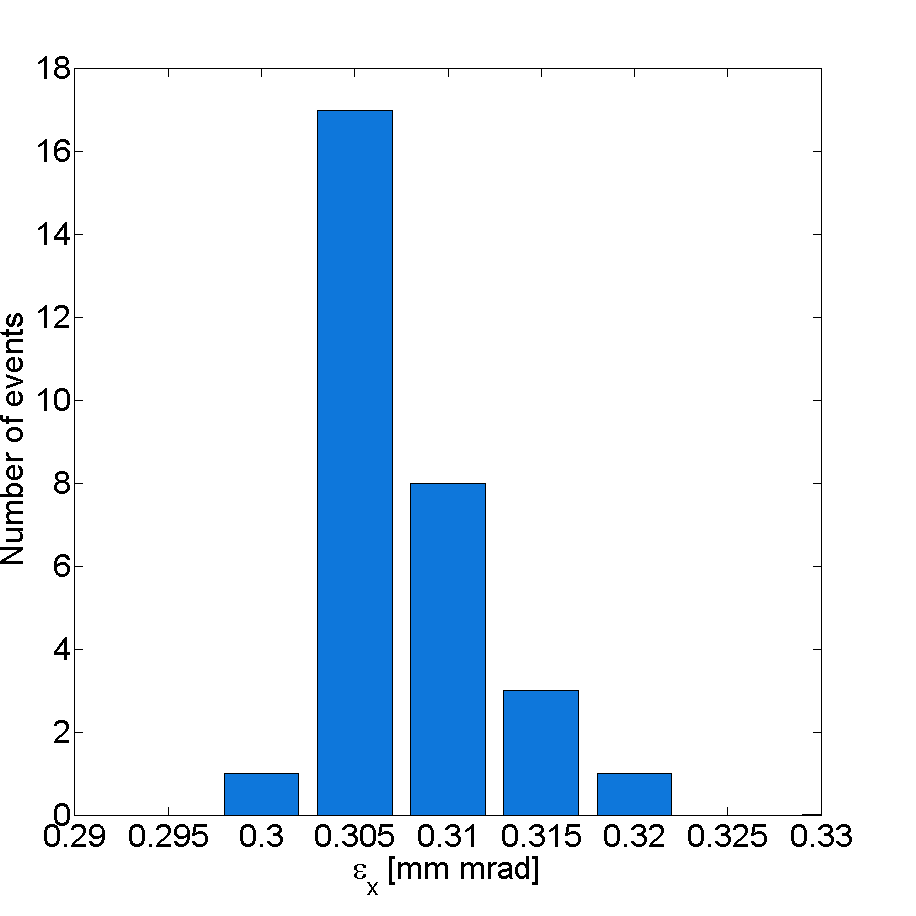}
     \caption{Emittance jitter distribution of 30 machines in the latin hypercube sampling analysis performed for the considered working point.}
     \end{figure}

     \begin{figure}[h!]
     \centering
     \includegraphics [width=.71\linewidth]{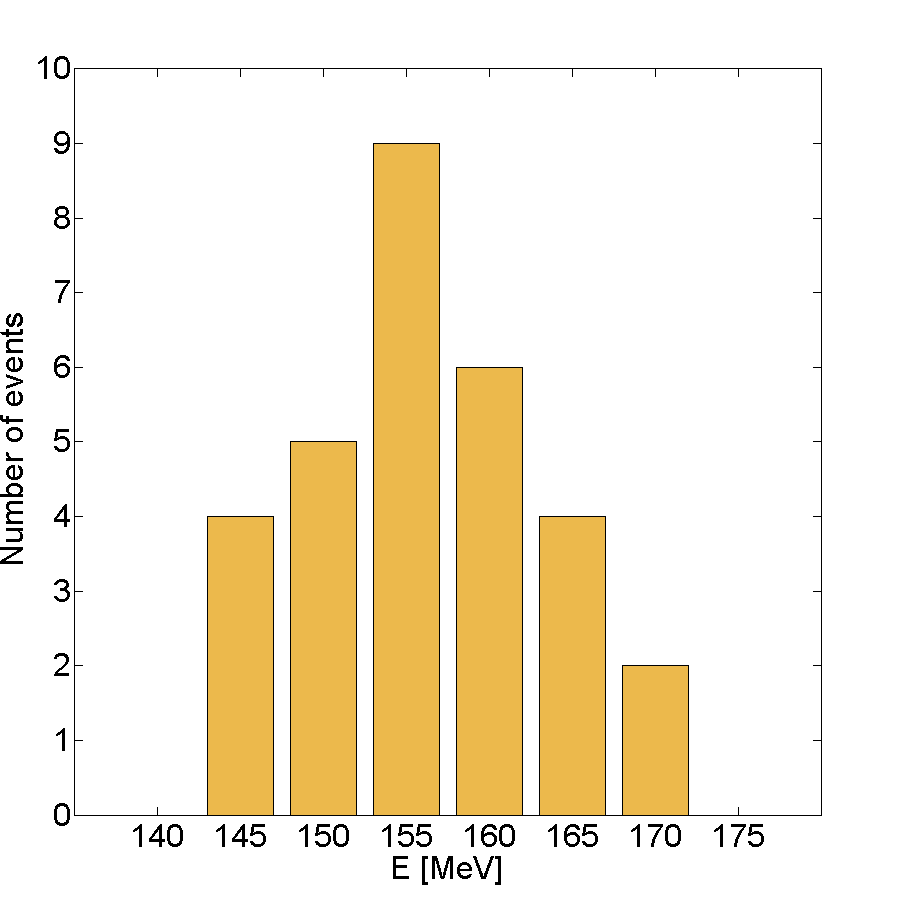}
     \caption{Energy jitter distribution of 30 machines in the latin hypercube sampling analysis performed for the considered working point.}
     \end{figure}     

\bibliographystyle{elsarticle-num}

\bibliography{biblio}

\end{document}